\documentclass[preprint,showpacs,preprintnumbers,amsmath,amssymb,superscriptaddress]{revtex4}

\usepackage{graphicx}
\usepackage{dcolumn}
\usepackage{bm}
\usepackage{pstricks}
\usepackage{epsfig}
\usepackage{psfrag}

\begin{document}
\title{Charge state control in single InAs/GaAs quantum dots by external
electric and magnetic fields}

\author{Jing Tang}
\affiliation{Institute of Photo-electronic Thin Film Devices and
Technology, Nankai University, Tianjin, 300071, China}
\affiliation{Beijing National Laboratory for Condensed Matter
Physics, Institute of Physics, Chinese Academy of Sciences, Beijing,
100190, China}

\author{Shuo Cao}
\author{Yunan Gao}
\author{Yue Sun}
\affiliation{Beijing National Laboratory for Condensed Matter
Physics, Institute of Physics, Chinese Academy of Sciences, Beijing,
100190, China}

\author{Weidong Geng}
\email{gengwd@nankai.edu.cn}
\affiliation{Institute of
Photo-electronic Thin Film Devices and Technology, Nankai
University, Tianjin, 300071, China}

\author{David A. Williams}
\affiliation{Hitachi Cambridge Laboratory, Cavendish Laboratory,
Cambridge CB3 0HE, United Kingdom}

\author{Kuijuan Jin}
\affiliation{Beijing National Laboratory for Condensed Matter
Physics, Institute of Physics, Chinese Academy of Sciences, Beijing,
100190, China}

\author{Xiulai Xu}
\email{xlxu@iphy.ac.cn}
\affiliation{Beijing National Laboratory for
Condensed Matter Physics, Institute of Physics, Chinese Academy of
Sciences, Beijing, 100190, China}

\date{\today}

\begin{abstract}
We report a photoluminescence (PL) spectroscopy study of charge state control in single self-assembled InAs/GaAs quantum dots by applying electric and/or magnetic fields at 4.2 K. Neutral and charged exciton complexes were observed under applied bias voltages from -0.5 V to 0.5 V by controlling the carrier tunneling. The highly negatively charged exciton emission becomes stronger with increasing pumping power, arising from the fact that electrons have a smaller effective mass than holes and are more easily captured by the quantum dots. The integrated PL intensity of negatively charged excitons is affected significantly by a magnetic field applied along the sample growth axis. This observation is explained by a reduction in the electron drift velocity caused by an applied magnetic field, which increases the probability of non-resonantly excited electrons being trapped by localized potentials at the wetting layer interface, and results in fewer electrons distributed in the quantum dots. The hole drift velocity is also affected by the magnetic field, but it is much weaker.

\end{abstract}


\maketitle
Self-assembled InAs/GaAs quantum dots (QDs) have been investigated
intensively to implement quantum information and quantum
computation, such as, single-photon sources~\cite{Kim1999,
Michler2000, Xu2004, Bennett2005,Xu2007}, exciton qubits and spin
qubits~\cite{Kolodka2007,Weiss2013,Zrenner2002}, quantum logic
gates~\cite{Bonadeo1998,Troiani2000, Besombes2003, Fei2012},
spin-photon entanglement
interfaces~\cite{Greve2012,Schaibley2013,Webster2014}, and quantum
memories~\cite{Lundstrom1999}. Among those investigations, single
spins in single dots are of particular interest because of their
relatively long coherence time at low
temperature~\cite{Warburton2013,Weiss2013}. To achieve single
charge or spin states, a precise charge control with certain spin in single quantum dots is
on demand. Charges in single InAs quantum
dots can be injected by resonant~\cite{Mete2007,He2013} or
non-resonant~\cite{Mar2011,Ediger2005} optical pumping, or by electrical pumping~\cite{Xu2004}, and then they can be manipulated by applying an electric field~\cite{Mar2011,Ediger2007} or an
external magnetic field~\cite{Moskalenko2009,Moskalenko2008} across
the quantum dots.

In an
\emph{n-i}-Schottky diode structure, the charge states of sandwiched quantum dots can
be controlled precisely using external
bias~\cite{Mar20111,Mar20112,Baier2001,Ediger2005,Bracker2005}.
For example, a quantum dot charging from +6e to -6e has been
demonstrated~\cite{Ediger2007} using photoluminescence (PL) spectroscopy. Using
magneto-photoluminescence spectroscopy, fine-structure
splitting~\cite{Bayer2002}, Zeeman
splitting~\cite{Toft2007,Hogele2005} and diamagnetic
shift~\cite{Schulhauser2002,Tsai2008} have been investigated for
different charge states in single quantum dots. Recently, coupling
between triply negatively charged states and the wetting layer
has also been shown using
magneto-photoluminescence~\cite{Van2013,Karrai2004}. Up to now,  only a few works have been reported on the charge state control by applying an external magnetic field~\cite{Moskalenko2009,Moskalenko2008}.
Moskalenko and colleagues~\cite{Moskalenko2008} reported a charge
redistribution in single quantum dots with increasing magnetic field
parallel to the sample growth direction (Faraday geometry), and
the redistribution was explained by the reduction of electron and hole
drifting velocities in the magnetic
field. In this Letter, we report a comprehensive
micro-PL study of the charge state control in single quantum dots
with varying pumping power, applying external electric field and
magnetic field. The charging states from the triply negatively charged
exciton ($X^{3-}$) to the singly positively charged exciton ($X^{+}$)
can be precisely controlled. At certain bias voltages, the intensities of the charged excitonic states can be modified with increasing magnetic field in Faraday geometry. In particular, the PL intensities of triply and doubly negatively charged excitons decrease
with increasing magnetic field, while that of the singly negatively charged exciton increases.

A schematic \emph{n-i}-Schottky structure is shown in Fig.1(a). A Si $\delta$-doped GaAs layer with a doping density
N$_{d}$ = 5$\times$ 10$^{12}$ cm$^{-2}$ is used as \emph{n} type region and contacted by annealed AuGeNi alloy.  At the active regime, a semitransparent Ti with a thickness
of 10 nm was used as Schottky contact, and on top an Al mask
with different apertures of 1-3 $\mu$m in diameter was patterned for addressing
single quantum dots. The device was mounted on an $\emph{xyz}$ piezoelectric stage and placed in a helium
gas exchange cryostat equipped with a superconducting magnet. The magnetic field was applied in Faraday geometry and the measurements were carried out at 4.2 K. In the micro-PL measurements, a semiconductor laser with a
wavelength of 650 nm was used as pumping source and was focused on
one of the apertures by a microscope objective with a large
numerical aperture of 0.83. The PL from single quantum dots was
collected with the same objective and dispersed through a 0.55 m
spectrometer, and the spectrum was detected with a liquid nitrogen
cooled charge coupled device camera with a spectral resolution of 60
$\mu$eV.

The PL measurements were firstly performed under bias voltages from -0.5 V
to 0.5 V. The neutral and charged exciton complexes were
observed in several individual dots with similar features. Figure
1(b) shows typical PL spectra of a single quantum dot (QD$_{-}$1)
as a function of bias voltage, where five discrete PL peaks can be
observed. Figure 1(c) presents the spectra at three bias
voltages of -0.5 V, 0 V and 0.5 V. The five emission lines in Figure
1(b) from top to bottom are assigned to the recombination from the singly
positively charged exciton (X$^{+}$), the neutral exciton (X$^{0}$), the singly, doubly and triply negatively charged excitons
(X$^{-}$, X$^{2-}$, X$^{3-}$), as labeled in the
figure. The assignment is essentially based on the analysis of the bias voltage and the
pumping power dependent PL intensities and peak positions.

With a non-resonant optical pumping, the optically generated electrons/holes relax to quantum
dots via the wetting layer as shown in the Figure 1(d). The electron
drift velocity is larger than that of hole because of the smaller
effective mass of electron. This induces that electrons are more
quickly being captured by the quantum dot than holes, resulting in that the quantum dots are more negatively charged without external field. In a Schottky diode structure as illustrated in Figure 1(d), both conduction and valence bands can be tuned with an external bias, which can be used to control the charging of the quantum dots. With a negative bias voltage, photo-generated electrons are more
easily to tunnel out and leave the quantum dot positively charged,
while for positive bias more photo-generated electrons are localized in
quantum dots. Therefore, the PL intensities from positively charged
excitons would decrease with increasing bias voltage, resulting in relatively stronger PL intensities from the negatively charged excitons. Therefore, the five peaks from left to right in Figure 1(c) are attributed to the charge states from more negative to positive in the quantum dot.

The photon energy from quantum dots is mainly determined by the
electron and hole energy levels in the energy band, but it is also
slightly modified by Coulomb interactions between electron-electron,
electron-hole and hole-hole. Due to the Coulomb interaction, the energy
of positive trion (X$^{+}$) is several meV higher than that of
the neutral exciton (X$^{0}$), while the peak of negative trion (X$^{-}$)
red-shifts by several meV \cite{Warburton1998}. However,
X$^{2-}$ with an additional electron to X$^{-}$ only contributes a
weaker Coulomb interaction, as the third electron in the quantum
dot occupies the p-shell which has a smaller wave-function overlap
with electrons and the hole in the s-shell. Hence, from the energy positions in QD$_{-}$1, we can firmly assign that the second emission line from the top in Figure
1(b) with an energy of 1.273 eV is the neutral exciton ($X^{0}$) emission
and the rest are as labeled. The binding energy between
$X^{2-}$ and $X^{-}$ ($X^{3-}$ and $X^{2-}$) is about 0.5 meV (1.7
meV) in this quantum dot. The coexist of the emission peaks of different
charging states at certain voltages (as marked by dashed lines) is
due to the thick tunneling GaAs barrier induced weak coupling between the 2DEG regime and the
quantum dots, where the thickness of the barrier is 50 nm in
this structure\cite{Mar2011,Kleemans2010}.

The charging process in quantum dots can be further confirmed with
an excitation power dependent PL mapping. Figure 2 shows the PL spectra
of QD$_{-}$1 as a function of bias voltage with different pumping
powers. At a low pumping power of 0.06 $\mu$W and at zero bias
voltage, only a single electron and a single hole can be generated and form a neutral exciton in a time as marked in the top panel. This confirms the neutral exciton emission assignment above. The
negatively charged exciton emission can only be identified with a
positive bias towards 0.5 V. With increasing pumping power to 0.53
$\mu$W and 1.46 $\mu$W, the negatively charged exciton peaks
($X^{-}$ and $X^{2-}$) become stronger and a negative bias voltage
is required to observe $X^{0}$ and $X^{+}$ emission. With a
pumping power at 2.37 $\mu$W, the $X^{-}$ and $X^{2-}$ peak
intensities are getting even stronger and $X^{3-}$ peak appears. The
negative bias voltage required to observe $X^{0}$ and $X^{+}$ peaks
is out of the applied voltage range. The trend that the higher the
pumping power, the more negatively charging in the quantum dots is
resulted from the fact that the electrons are more easily being
captured than holes due to its smaller effective mass.

In the above, we show that the charge states in the quantum dot can be controlled by applying a bias voltage, where the carrier capture process plays an important role. In the following work, we show the charge state control by applying magnetic field. In the experiments, the excitation laser energy (1.908 eV) is higher than
the energy gap of GaAs barrier (1.518 eV). Hence, the optically
generated electrons and holes with certain kinetic energies will
firstly emerge at GaAs matrix separately or in pairs, then relax to
the wetting layer and finally be captured by the quantum
dots~\cite{Fry2000}. The applied bias modifies the band structure of
the devices as shown in Figure 1(d), which mainly affects the
capture process along the growth direction. The laser spot
size (1$\sim$2 $\mu$m in diameter) is far larger than the dot size
($\sim$25 nm), the drifting of optically generated carriers in
quantum dot plane cannot be avoided and provides an additional
way to control the charging states of quantum dots. Applying a
magnetic field perpendicular to the device plane (Faraday geometry)
can affect the electron and hole in plane transport by magnetic
field induced cyclotron motion~\cite{Moskalenko2008}.

Figure 3(a) shows the PL spectra under magnetic fields varied from 0 T to 9 T,
and in each panel the bias voltage varies from -0.5 V to 0.5 V. All
the spectral lines from different charged states (as marked on top of
the figure) split with increasing magnetic field, which is due to Zeeman effect. The similar \emph{g} factors for all the charged
states are calculated to be about 2.5$\pm0.08$, similar to the
reported values~\cite{Tsai2008}. The average peak position of the
splitting peaks shifts towards the high energy with increasing
magnetic field, which is due to the diamagnetic shift. It can be seen that the
integrated intensities of $X^{3-}$ and $X^{2-}$ decrease gradually
with an increase of applied magnetic field, and the $X^{3-}$ peak almost
quenched with a magnetic field at 6 T, while that of $X^{-}$
increases. In addition, the intensities of $X^{0}$ and $X^{+}$
decrease slightly with increasing magnetic field.

For clarity, Figure 3(b) shows the PL spectra in the negatively charged exciton
regime as a function of magnetic field with bias voltages at -0.5
V, 0 V and +0.5 V. In each panel, the PL intensity of $X^{-}$ increases with increasing magnetic
field, while the PL intensities of $X^{2-}$ and $X^{3-}$ decrease.
To obtain a clear intensity variation induced by the applied
magnetic field, the integrated intensities of the negatively charged
excitons for each branch of the Zeeman splitting ($\sigma^+$ and
$\sigma^-$) are plotted in Figure 3(c) at three bias voltages.
At zero magnetic field with a bias voltage at -0.5 V, only $X^{-}$ and $X^{2-}$ can be
observed with relatively low intensities, and the $X^{-}$ and
$X^{2-}$ are getting stronger and $X^{3-}$ appears at bias voltages
of 0 V and 0.5 V. At 0.5 V, the intensities of $X^{2-}$ and $X^{3-}$
decrease with applying magnetic field and $X^{3-}$ intensity approaches to zero with a magnetic field up to 6 T. However, $X^{-}$ peak is very weak at 0 T and quickly increases with a
magnetic field tuned to 4 T and then saturates. We ascribe this
charging state control to that the carrier
transport in quantum dot plane is modified by the magnetic field.

Due to the growth induced variations of alloy composition and strain
in the quantum dot plane, localized potential fluctuations can be formed
at the InAs/GaAs interfaces~\cite{Lobo1999,Krzyzewsk2002,cao2013}.
The potential fluctuations could trap the optically excited electrons and holes, and affect the
carrier relaxation, which ultimately affects the PL
intensities of the quantum dots~\cite{Lobo1999, Toda1999}. Owing to that the
effective mass of hole ($m_h=0.45~m_0$) is $\sim$6.72 times larger than
that of electrons ($m_e=0.067~m_0$), where $m_0$ represents the
mass of free-electron~\cite{Hillmer1989}, holes are more easily being
trapped by the localized potentials than electrons. This induces
a larger probability for electrons being captured by quantum dots
than holes. It could be an additional reason for that the
negatively charged exciton peaks are getting stronger with
increasing pumping power without applied electric and magnetic
fields (as shown in Figure 2).

Moskalenko et al.~\cite{Moskalenko2008} recently have studied the
influence of the vertical magnetic field on the electron and hole
transport in similar quantum dots, where they found that the PL intensity of
$X^{0}$ increases while those of $X^{-}$ and $X^{2-}$ decrease with
increasing magnetic field. The magnetic field applied in Faraday
geometry induces the cyclotron motion of electrons and holes, which
reduces their drifting velocities. The reduced drifting of electrons
and holes would raise their probability of being trapped by local
potential prior to arriving at the quantum dot. This model requires
that $\omega_{c}^{e(h)}\times\tau_{sc}^{e(h)}$ is larger than
one~\cite{Moskalenko2008,Kittle1976}. Here $\omega_{c}^{e(h)}$ is the
cyclotron frequency $e^*B/m_{e(h)}$, where $e^*$ is the elementary
charge, $\tau_{sc}^{e(h)}$ is the scattering time of an electron
(hole). We used the scattering time of 3.4 ps and 0.74 ps for
electrons and holes in GaAs from reference~\cite{Moskalenko2008,Bockelmann1990}. To achieve
$\omega_{c}^{e(h)}\times\tau_{sc}^{e(h)}>1$, the minimum magnetic
field needed for affecting electrons (holes) is 0.11 T (3.4 T). It indicates that electrons are much easier than holes to be affected by the
magnetic field. This means that more electrons would be affected
during the drifting in the quantum dot plane and trapped by the localized
potentials at 4.2 K~\cite{Moskalenko2008,cao2013}, resulting in a
reduced negatively charging in quantum dots with increasing magnetic
field. Therefore the PL intensities of $X^{3-}$ and $X^{2-}$
decrease while that of $X^{-}$ increases. The magnetic
field was not strong enough to observe the intensity increases of $X^{0}$ and $X^{+}$ yet. On the
contrast, a slight intensity decrease of $X^{0}$ and
$X^{+}$ could be due to the increased number of holes being
trapped by the localized potentials in high magnetic field.

To further examine the model and show the reproducibility of the
observation, Figure 4 shows the PL spectra of another quantum dot
(QD$_{-}$2) as a function of magnetic field at different bias
voltages. The charged exciton peaks are labeled and the dashed lines mark the Zeeman
splitting. Similar features as
QD$_{-}$1 can be observed. At -0.5 V, $X^{-}$ peak (as labeled in
the Figure 4) dominates in the spectra and total intensity is
relatively low. As the bias voltage increases, the $X^{2-}$
intensity increases and dominates in the spectrum at 0 T. However,
the intensity of $X^{2-}$ quickly decreases while that of $X^{-}$
increases with a magnetic field only at about 2 T, which further supports above discussion.

In conclusion, charging state control has been demonstrated in
single quantum dots at 4.2 K by varying pumping power, applying
an external electric field and/or a magnetic field. Because the effective
 mass of electron in GaAs is much less than that of hole, electrons have
a larger drift velocity and a larger probability being captured by
quantum dots, resulting in stronger PL intensities of negatively
charged excitons with increasing the pumping power. In addition, the recombination from different
charged states (from $X^{3-}$ to $X^{+}$) can be precisely
controlled by an applied electric field. With a magnetic field
applied parallel to the growth direction, electron
transport is more easily to be affected by the magnetic field than
that of hole, because the electrons can form cyclotron orbit with a
smaller magnetic field. This increases the probability of the
electrons being trapped by the localized potentials in the quantum
dot plane, which results in carrier redistribution in quantum dots.
Our results suggest that by tuning electric field, magnetic field
and pumping power, the charging states of a single quantum dot
can be precisely controlled.

\section{Acknowledgements}

This work was supported by the National Basic Research Program of
China under Grant No. 2013CB328706 and 2014CB921003; the National
Natural Science Foundation of China under Grant No. 11174356 and
61275060; the Strategic Priority Research Program of the Chinese
Academy of Sciences under Grant No. XDB07030200; the Hundred Talents
Program of the Chinese Academy of Sciences, and the China
Postdoctoral Science Foundation under Grand No. 2013M540155. We
thank Andrew Ramsay, Jonathan Mar, Weidong Sheng and Mete
Atat\"{u}re for very helpful discussions.

\newpage
\begin{figure*}
\centering

\epsfig{file=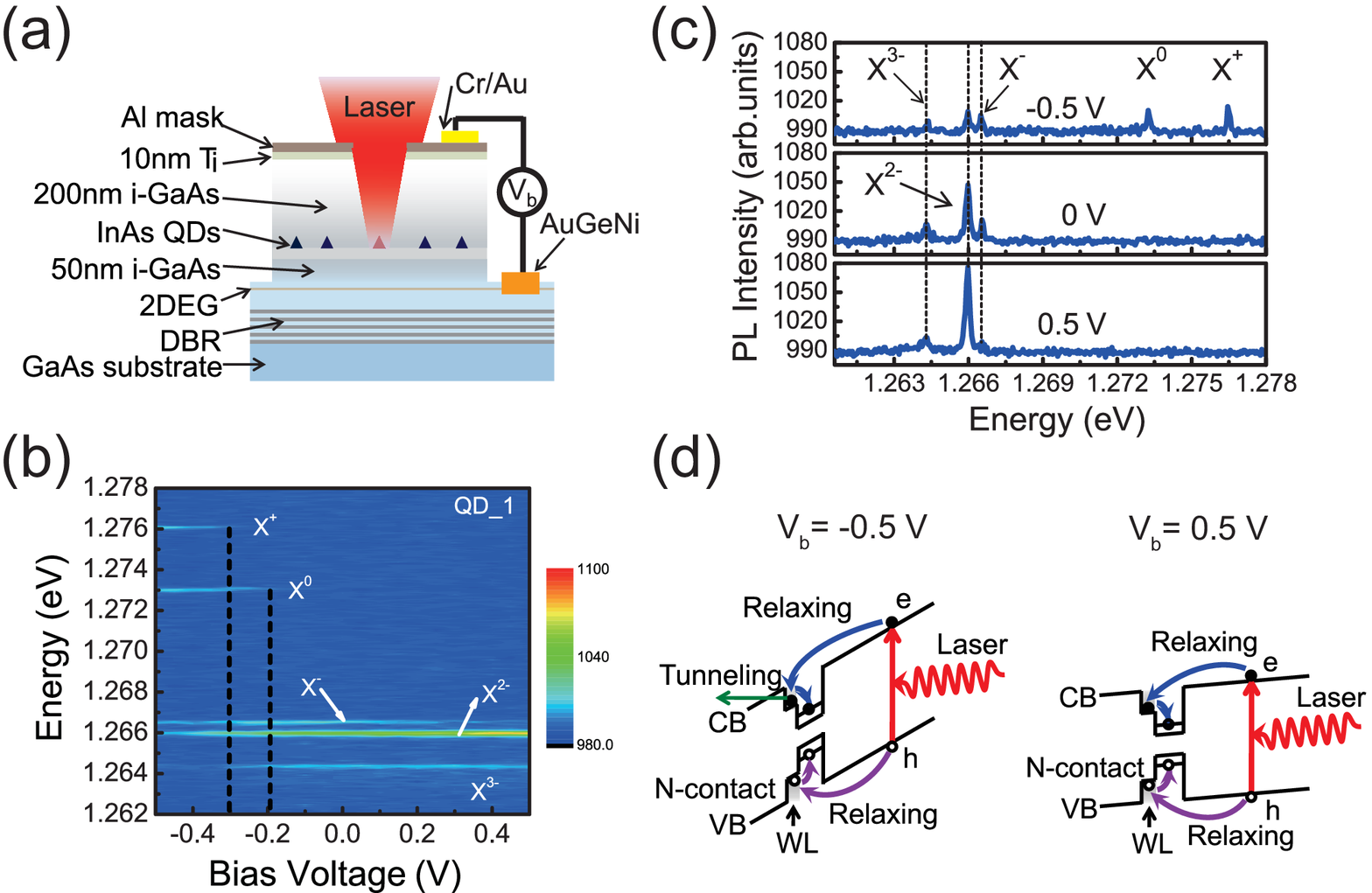,width=12cm,keepaspectratio}
\caption{\label{fig:figure1} (a) A schematic diagram of a \emph{n-i-}Schottky device. (b) PL spectra with bias voltages sweeping from -0.5 V to +0.5 V of QD$_{-}$1 with an
excitation power of 2.37 $\mu$W. The PL peaks from
$X^{3-}$,~$X^{2-}$,~$X^{-}$,~$X^{0}$ and $X^{+}$ are labeled in the
figure. The dotted lines mark the beginning of emission lines of
~$X^{0}$ and $X^{+}$. (c) PL spectra with bias voltages at -0.5 V,
0 V and +0.5 V from the top to the bottom panels. (d) Band profiles of the
\emph{n-i-}Schottky diode structure under bias voltages of -0.5 V
and +0.5 V.}
\vspace{5cm}
\end{figure*}

\newpage
\begin{figure*}
\centering
\epsfig{file=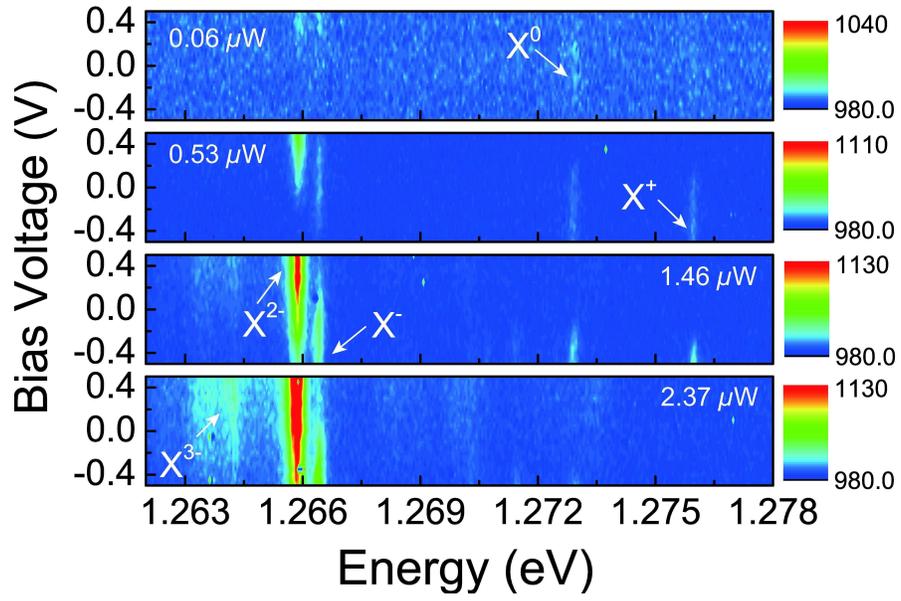,width=12cm,keepaspectratio}
\caption{\label{fig:figure2} PL spectra of QD$_{-}$1 as a function of bias voltage from -0.5 V to
+0.5 V with different pumping powers. The pumping powers from the
top panel to the bottom panel are 0.06 $\mu$W, 0.53 $\mu$W, 1.46
$\mu$W and 2.37 $\mu$W, respectively.}
\vspace{5cm}
\end{figure*}

\newpage
\begin{figure*}
\centering
\epsfig{file=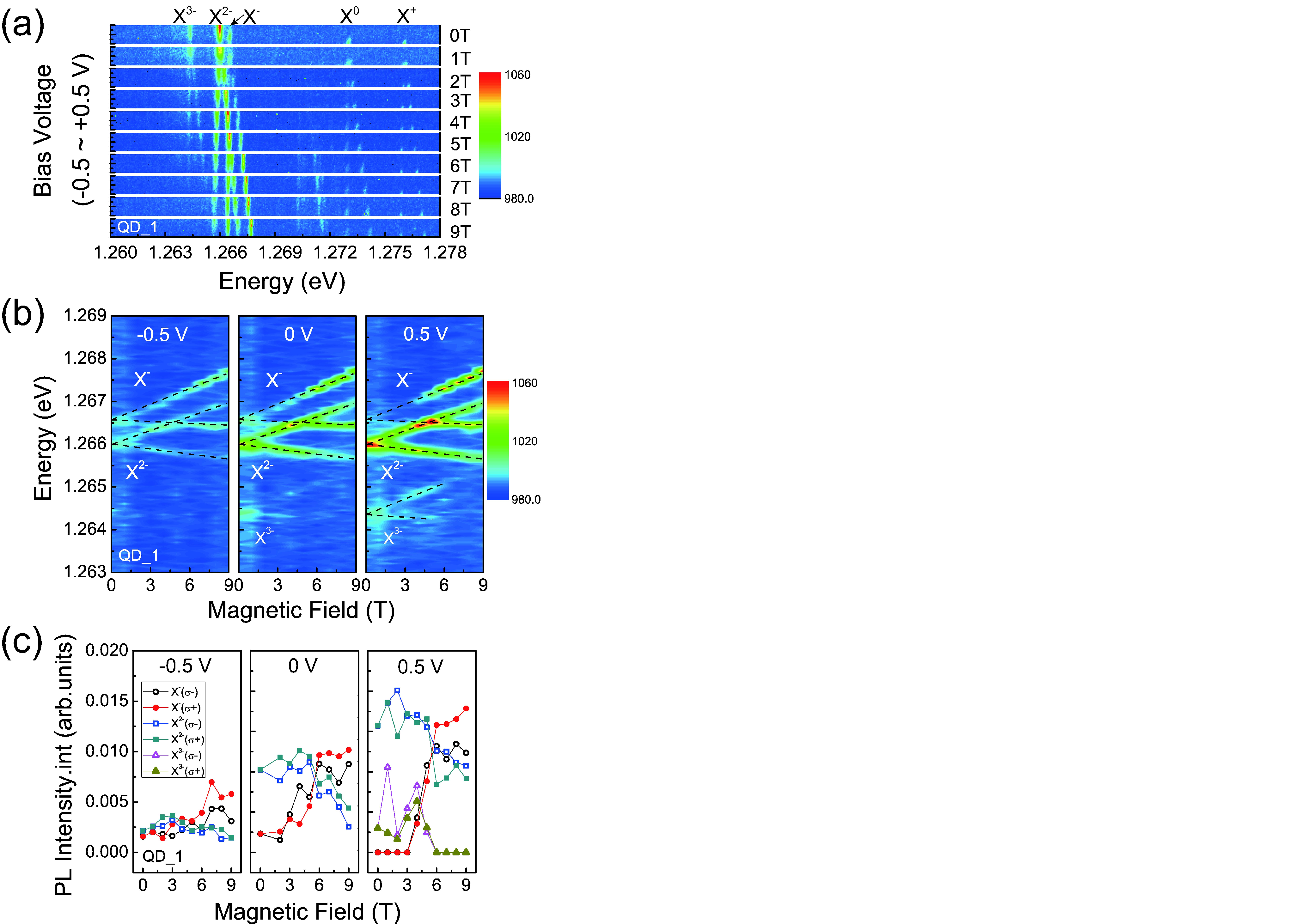,width=10cm,keepaspectratio}
\caption{\label{fig:figure3} (a) Contour plot of the PL spectra of QD$_{-}$1 as a function of
bias voltage from -0.5 V to +0.5 V with different magnetic fields
from 0 T to 9 T in Faraday geometry. Each panel shows PL spectra of
QD$_{-}$1 as a function of bias voltage from -0.5 V (bottom) to +0.5
V (top). (b) PL spectra of $X^{-}$, $X^{2-}$ and $X^{3-}$ in
QD$_{-}$1 as a function of applied magnetic field from 0 T to 9 T
with bias voltage at -0.5 V, 0 V and +0.5 V, respectively. The
dashed lines mark the corresponding energy levels by the Zeeman
splitting. (c) The integrated intensities of $X^{-}$, $X^{2-}$ and
$X^{3-}$ as a function of the applied magnetic field at different bias voltages.}
\vspace{5cm}
\end{figure*}

\newpage
\begin{figure*}
\centering
\epsfig{file=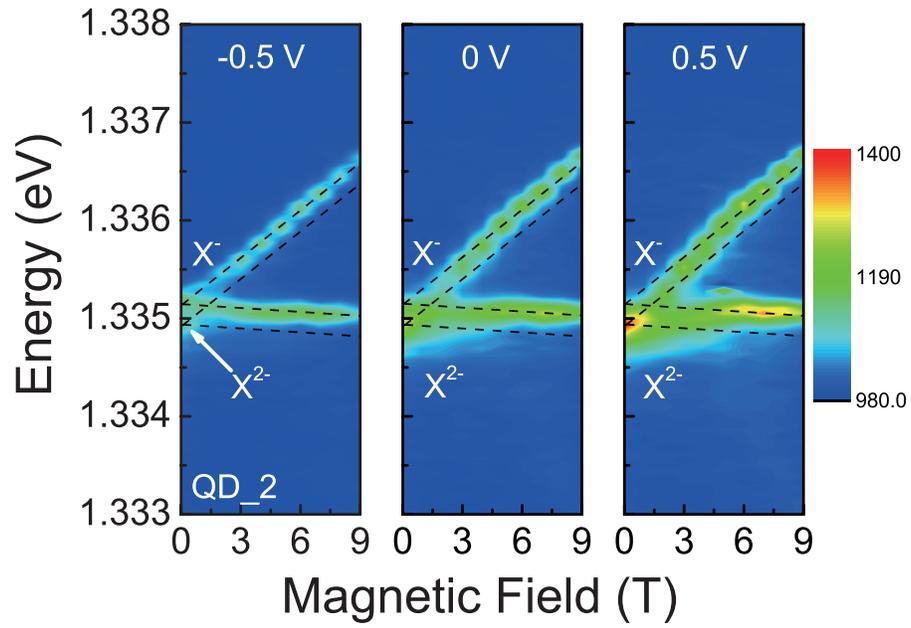,width=12cm,keepaspectratio}
\caption{\label{fig:figure4} PL spectra of $X^{2-}$ and $X^{-}$ from QD$_{-}$2 as a function of
applied magnetic field from 0 T to 9 T, with bias voltages at -0.5
V, 0 V and +0.5 V. The dashed lines mark the corresponding energy
levels by the Zeeman splitting for different charge states.}
\vspace{5cm}
\end{figure*}

\end{document}